\documentclass[12pt,preprint]{aastex}

\usepackage{amsmath}

\shorttitle{Metallicities in SFH Measurements}
\shortauthors{Dolphin}

\begin{document}


\title{On the Incorporation of Metallicity Data into Star Formation History Measurements from Resolved Stellar Populations}


\author{Andrew E. Dolphin}
\affil{Raytheon Company, Tucson, AZ, 85734}
\email{adolphin@raytheon.com}

\begin{abstract}
The combination of spectroscopic stellar metallicities and resolved star color-magnitude diagrams (CMDs) has the potential to constrain the entire star formation and chemical enrichment history (SFH) of a galaxy better than fitting CMDs alone (as is most common in SFH studies using resolved stellar populations).  In this paper, two approaches for incorporating external metallicity information into color-magnitude diagram fitting techniques are presented.  Overall, the joint fitting of metallicity and CMD information can increase the precision on measured age-metallicity relationships and star formation rates by ~10\% over CMD fitting alone.  However, systematics in stellar isochrones and mismatches between spectroscopic and photometric metallicity determinations can reduce the accuracy of the recovered SFHs.  I present a simple mitigation of these systematics that can reduce the amplitude of these systematics to the level obtained from CMD fitting alone, while ensuring the age-metallicity relationship is consistent with spectroscopic metallicities.  As is the case in CMD-fitting analysis, improved stellar models and calibrations between spectroscopic and photometric metallicities are currently the primary impediment to gains in SFH precision from jointly fitting stellar metallicities and CMDs.
\end{abstract}





\section{Introduction \label{sec-intro}}

The use of color-magnitude diagrams (CMDs) of nearby galaxies with resolved stellar populations has provided a wealth of understanding of the star formation histories (SFHs) of these systems.  At a high level, the relative population of age-sensitive features such as the upper main sequence and red giant branch indicate the relative star formation rates at young vs. old ages.  Additionally, features such as the red giant branch color are indicative of metallicity.  Thus, an examination of all age- and metallicity-sensitive CMD features can be used to infer the qualitative SFHs of these galaxies, which \citet{hod89} depicted in the form of population boxes (star formation rate vs. lookback time and metallicity).  Note that throughout this paper, the SFH is defined as a combination of the star formation rate (SFR) and age-metallicity relationship (AMR).

Quantitative approaches to this problem have been developed more recently \citep[e.g.,][]{gal96,tol96,dol97,her99,hol99,har01}, in which synthetic CMDs are simulated based on trial SFHs, and the quality of the fit between the observed data and a synthetic CMD is used to infer the correctness of the trial SFH.  While these methods differ in terms of parameterization and CMD binning, the key similarity is that all use a quantitative metric to measure the goodness of the CMD fit.

In addition to measuring the number of stars formed as a function of lookback time, CMD fitting also provides information regarding the metallicity distribution.  The most common use of photometry to infer metallicity is in the red giant branch \citep[e.g.,][]{dac90}, though most features on the CMD have sufficient sensitivity to metallicity that population boxes can be estimated from the CMDs alone \citep[e.g.,][]{dol05}.

While CMD-fitting techniques are capable of estimating the age-metallicity relationship, inclusion of spectroscopic metallicity measurements in the solutions \citep{deb14} provides an independent, direct constraint.  In addition to providing improved constraints on the AMR, this has been shown to provide a better constraint to the star formation rates estimated from CMD features with age-metallicity degeneracies \citep[e.g.,][]{wor94}.  Since CMD-fitting techniques are inherently self-consistent (accurately measuring SFHs of synthetic populations created from the same models), this increased precision is potentially obtained without sacrificing accuracy.

The presence of increased constraints on the solution, however, must be examined to ensure the resulting solution is robust.  As demonstrated by \citet{dol12}, even CMD-only solutions can be sufficiently constrained that the presence of systematic errors (e.g., differences between underlying stellar models) produces a solution that is inconsistent with the input SFH.  Incorporating metallicity information into a solution may compound this discrepancy.  For example, consider the case in which the metallicity of a real star has been measured, but that no combination of other physical parameters (e.g., age, extinction), given this metallicity, cause an isochrone to overlap the star's CMD location.

This analysis seeks to evaluate the potential for improving SFH estimates by inclusion of metallicity information in a CMD solution, assess the magnitude of errors that can be created by systematic uncertainties, and propose a robust approach to minimize such errors.

\section{Incorporation of Metallicity Data \label{sec-simple}}

For stars for which both metallicity measurements and photometry are available, the CMD fitting technique commonly used for SFH measurements can be extended by adding a third dimension to the solution, resulting in a color-magnitude-metallicity diagram (CMMD).  However, this is likely to result in a large amount of data discarded, as the photometric and spectroscopic samples are unlikely to consist of the same stars.  Both conditions (stars in only photometric or only spectroscopic data) must be considered.

Stars present only in the photometric sample are the simpler case.  These are often dim stars that have adequate signal-to-noise for photometric measurements but not for spectroscopic metallicity determinations.  For example, many nearby galaxies have HST photometry that is significantly deeper than the spectroscopic samples, such as Milky Way companions with photometry reaching the ancient main sequence turnoff but stellar metallicities only for red giants.  To account for these stars, an extra metallicity bin can be added to the CMMD to include those stars with photometric data only.  In order to create the synthetic CMMD, one must also define a metallicity completeness function that provides the probability of a star having a metallicity measurement.  For the examples shown in this study, this is a simple function of color and magnitude.

The other special case for consideration is stars for which metallicity measurements are available but photometric measurements are not.  This is common for systems with ground-based spectroscopy covering a much larger field than HST-based photometry.  In this case, a metallicity distribution function (MDF) for these stars can be fit simultaneously with the CMD.  In order to create the synthetic MDF, one must be able to quantify color and magnitude limits for the sample of stars and assume that the stars comprising the metallicity sample come from a population representative of those in the photometric sample.  Additionally, the CMD + MDF fitting technique must account for the MDF sample covering a different spatial region than the CMD sample; for the examples shown in this study the synthetic MDF is normalized to contain the same total number of stars as the observed MDF.

A final consideration in creating the synthetic metallicity estimates is measurement uncertainty.  While artificial star tests are commonly used for estimating photometric completeness and noise, an equivalent approach is not commonly used for metallicity measurements.  Thus, the software used in this study \citep[an extension to][]{dol02} incorporates a Gaussian error model whose width is a function of color, magnitude, and age.

\begin{figure}
\epsscale{1.0}
%\plotone{../simple/deep_old_sparse_b.ps}
\plotone{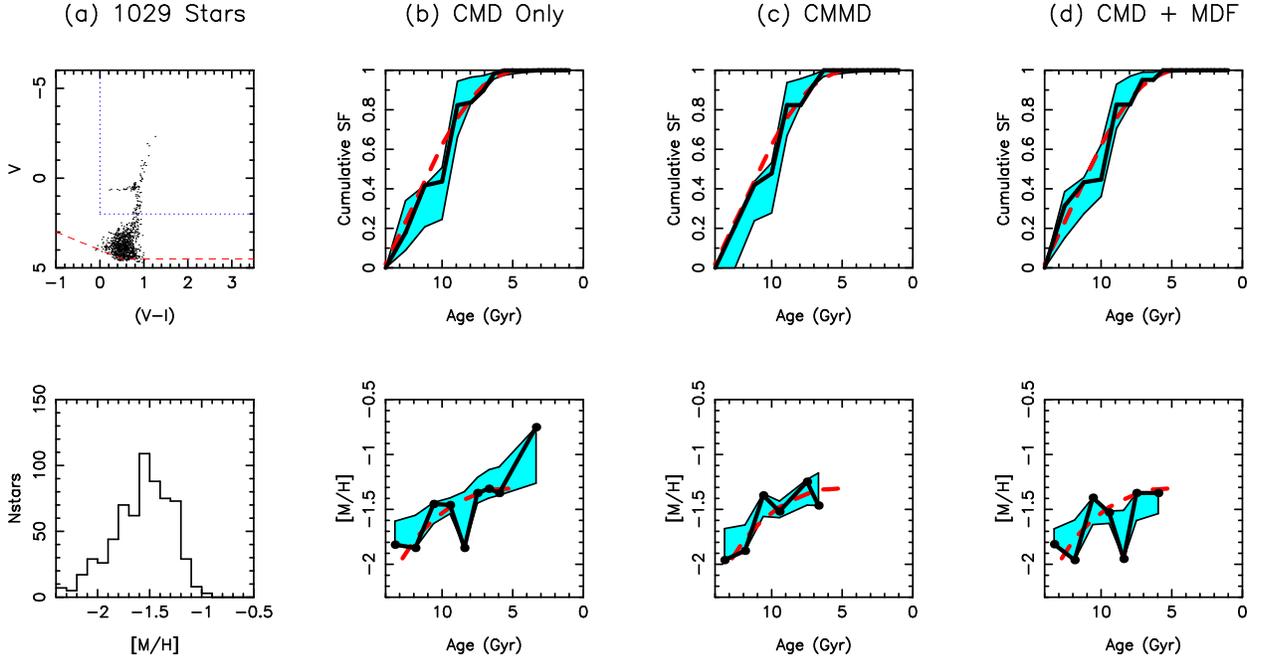}
\caption{The left panels (a) show the simulated photometry and MDF for an early-type galaxy with deep photometry (reaching the old main sequence turnoff).  The red dashed line in the CMD shows 50\% completeness limits; the blue dotted line shows the region from which the MDF was taken.  Note the MDF region contains only 10\% of the stars in the CMD, but metallicity data are from a field 6.5 times larger than the photometric field.  Panel (b) shows the SFR and AMR used to create the CMD (dashed red) and those obtained by fitting the CMD (solid black).  The cyan area shows the uncertainties due to random errors.  Panels (c) and (d) show the solutions for CMMD-fitting and CMD+MDF fitting, respectively.  In this example, the MDF is from a larger region than the photometric sample to illustrate an ideal case for a CMD+MDF solution.
\label{fig-example1}}
\end{figure}

\begin{figure}
\epsscale{1.0}
%\plotone{../simple/shallow_young_full.ps}
\plotone{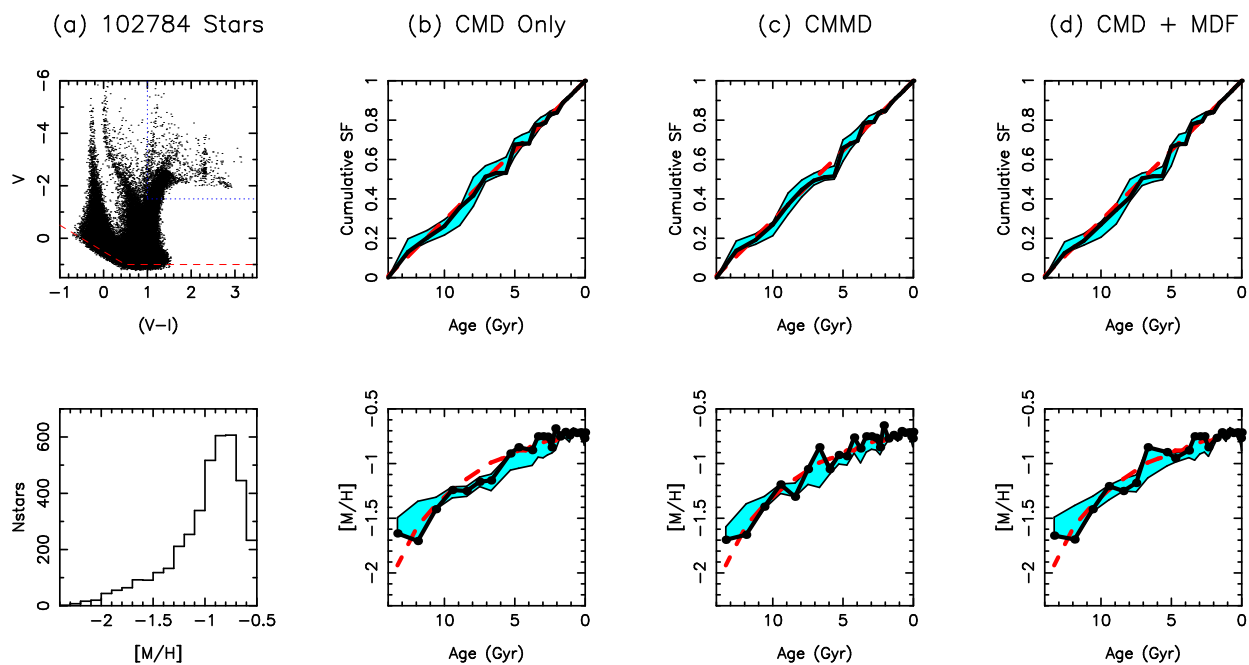}
\caption{Same as Figure \ref{fig-example1}, but for a late-type galaxy with shallow photometry (reaching the horizontal branch but not old main sequence turnoff).  In this example, the MDF is from the same stars with photometry (about 6\% of the total) to illustrate an ideal case for a CMMD solution.  Note that the number of stars is significantly more than in Figure \ref{fig-example1}, resulting in a factor of ~10 reduction in random errors.
\label{fig-example2}}
\end{figure}

To evaluate the potential improvement in SFH measurement constraints achieved by including metallicities, tests were run for a variety of depths, stellar populations, and number of stars in the photometric sample.  Two examples from these tests are shown in Figures \ref{fig-example1} and \ref{fig-example2}.  Both examples show SFH solutions with and without metallicity information, with the metallicity solutions improving the accuracy of not just the recovered AMR but also the SFR.
\placefigure{fig-example1}
\placefigure{fig-example2}

The population in Figure \ref{fig-example1} is intended to represent a nearby early-type galaxy for which the photometry was obtained from a smaller field of view than were the metallicities.  In this example, the star formation history (both SFR and AMR) error bars were reduced by $\sim 12\%$ when incorporating the MDF of the entire spectroscopic sample into the solution, while only reduced by $\sim 6\%$ when using just the metallicities of the stars with photometry.

In Figure \ref{fig-example2}, the population is intended to represent a more distant, late-type galaxy for which the photometry and metallicities both cover the entire system.  Here, $\sim 11\%$ reductions in the uncertainties were obtained in both metallicity solutions.

It is important to note that the SFH measurements presented in Figures \ref{fig-example1} and \ref{fig-example2} are for a best-case situation in which the stellar models are perfect.  Thus, these results confirm findings from \citep{dol02} showing that excellent SFR and AMR recovery is possible with CMDs alone, and thus that inclusion of metallicity data should not be expected to provide a significant improvement.

\section{Effects of Isochrone Uncertainties \label{sec-systematics}}

While the examples shown in Section \ref{sec-simple} provide encouragement that metallicities can be used to improve SFH solutions, those examples do not incorporate any effects of isochrone uncertainties (e.g., input physics, rotation, bolometric corrections).  While isochrone uncertainties are important to consider in CMD-fitting studies \citep{apa09,dol12}, it is generally possible to obtain a good CMD solution in the presence of isochrone mismatches by allowing for an offset in age or metallicity.  This is not necessarily the case in a CMMD solution, however, as it is possible that the isochrones corresponding to the measured metallicity may not fall at the CMD location of the measured photometry.  In this case, finding an acceptable fit may be impossible.

The solutions shown in Figures \ref{fig-example1} and \ref{fig-example2} did not include isochrone uncertainties, as the Padua models \citep{mar08,gir10} were used to both create the simulated data and measure the SFHs.  In contrast, Figure \ref{fig-systematic} shows the same simulated data, but this time using the PARSEC models \citep{bre12} to measure the SFHs.  Note that the only source of error introduced here is in the isochrones; other sources of error such as imperfections in photometric error or internal reddening models are not reflected.
\placefigure{fig-systematic}

Panel (a) of Figure \ref{fig-systematic} shows the systematic errors in the CMD solution, which are consistent with what was shown by \citet{dol12}.  From panels (b) and (c), however, one sees that incorporation of metallicity information creates larger errors in the measured star formation rates, in the sense that all star formation between 1.5 Gyr and 6 Gyr ago was pushed into a burst of age $\sim 1.5$ Gyr.  In effect, a CMD-only solution will tend to select isochrones with close to the correct age by using incorrect metallicities.  However, once metallicity constraints are added, this is not possible and the star formation rates will be in error.

In Figure \ref{fig-systematic}, the systematic error in the CMMD solution is also seen to be worse than that in the CMD-only and CMD+MDF solutions.  This is due to the fact that metallicities are assigned to specific stars on the CMD rather than generally to a MDF, making it more likely that stars cannot be fit by any isochrone in the set.  The goodness-of-fit measurements of the solutions also indicate that the CMMD solution suffered the most degradation from the systematic errors.

\section{Mitigation of Isochrone Uncertainties \label{sec-correction}}

Given the risk of CMMD (or CMD+MDF) solutions introducing large systematic errors, one must carefully consider how to incorporate metallicity information.  The approach proposed in this section is motivated by the observation in Section \ref{sec-systematics} that CMD-fitting will account for isochrone errors by adopting metallicities shifted from the true metallicities.

When creating synthetic metallicity estimates, the synthetic metallicity can be biased relative to the metallicity of the isochrones used to populate the CMDs.  That is, the metallicity error model does not need to have a mean of zero, but instead can have a mean that is a function of various parameters (the software used in this study allows variation with color, magnitude, and age).  Note that increasing the random component of the error model would also be appropriate, though is not done in the example shown here.

Using this approach, the synthetic data that resulted in a significant SFH error in Figure \ref{fig-systematic} were re-solved using an age-dependent metallicity offset.  Note that this offset was obtained empirically by minimizing the fit parameter, not by a more detailed comparison of the two isochrone sets (which would be possible for simulated data but not real data).  Results of this solution are shown in Figure \ref{fig-sys_fixed}.
\placefigure{fig-sys_fixed}

Because it is based on the CMD only, the left panel (a) of Figure \ref{fig-sys_fixed} is the same solution from Figure \ref{fig-systematic}, except with the age-dependent metallicity shift added.  The other two panels, however, show significant improvements in the SFH solutions, with both the metallicities and star formation rates much closer to the input (dashed line) than was the case in Figure \ref{fig-systematic}.  Specifically, the artificial 1.5 Gyr old burst was reduced by $\sim 25\%$, and the bias towards measuring stars at younger ages has been reduced.  By comparison with the CMD-only solution, it appears that the metallicity shifts have mitigated the additional systematic errors created by incorporating metallicity information into the solutions.  In effect, the result is that the measurement of SFR vs. time is largely unaffected, while the AMR has been shifted from an isochrone scale to a spectroscopic metallicity scale.

Despite having mitigated the additional error created by incorporating metallicity data, the measured star formation histories are still affected by systematic errors comparable to those from CMD-only solutions.  Furthermore, the SFR error bars in Figures \ref{fig-systematic} and \ref{fig-sys_fixed} are approximately half the size of those in Figure \ref{fig-example2}.  This is a result of systematic differences between the isochrone sets artificially constraining the solution space.  Thus, it is necessary to include systematic uncertainties when reporting star formation histories, as isochrone uncertainties affect both the measured SFH and the random uncertainties.

A final observation is that, despite incorporating knowledge of metallicities of specific stars in the CMD, the CMMD solution in Figure \ref{fig-sys_fixed} panel (b) is not significantly different from the CMD+MDF solution in panel (c).  Also considering that the CMMD solution requiring an order of magnitude more runtime and being more vulnerable to systematics (as seen in Figure \ref{fig-systematic}), the recommended method to incorporate metallicities into a SFH measurement is to fit an MDF along with the CMD solution rather than adding an extra dimension to create a CMMD.  Note that this is merely a separation of a CMMD into a CMD and an MDF (separating metallicity data from photometric data); modeling issues such as the selection model are unchanged.

\section{Summary \label{sec-summary}}

Incorporation of metallicity information into CMD-based SFH measurements has the potential to increase the constraints on the measured SFH, with both the SFR and AMR seeing comparable reduction in uncertainties.  In this study, a three-mode approach to ensure all data are incorporated is proposed: a single SFH solution using a CMMD for stars with photometry and metallicities, a CMD for stars with photometry only, and an MDF for stars with metallicities only.  This will enable optimal solutions to be made in the most common cases where the spectroscopic and photometric fields are of different depth or cover different fields of view.

\begin{figure}
\epsscale{1.0}
%\plotone{../models/shallow_young_full.nz.ps}
\plotone{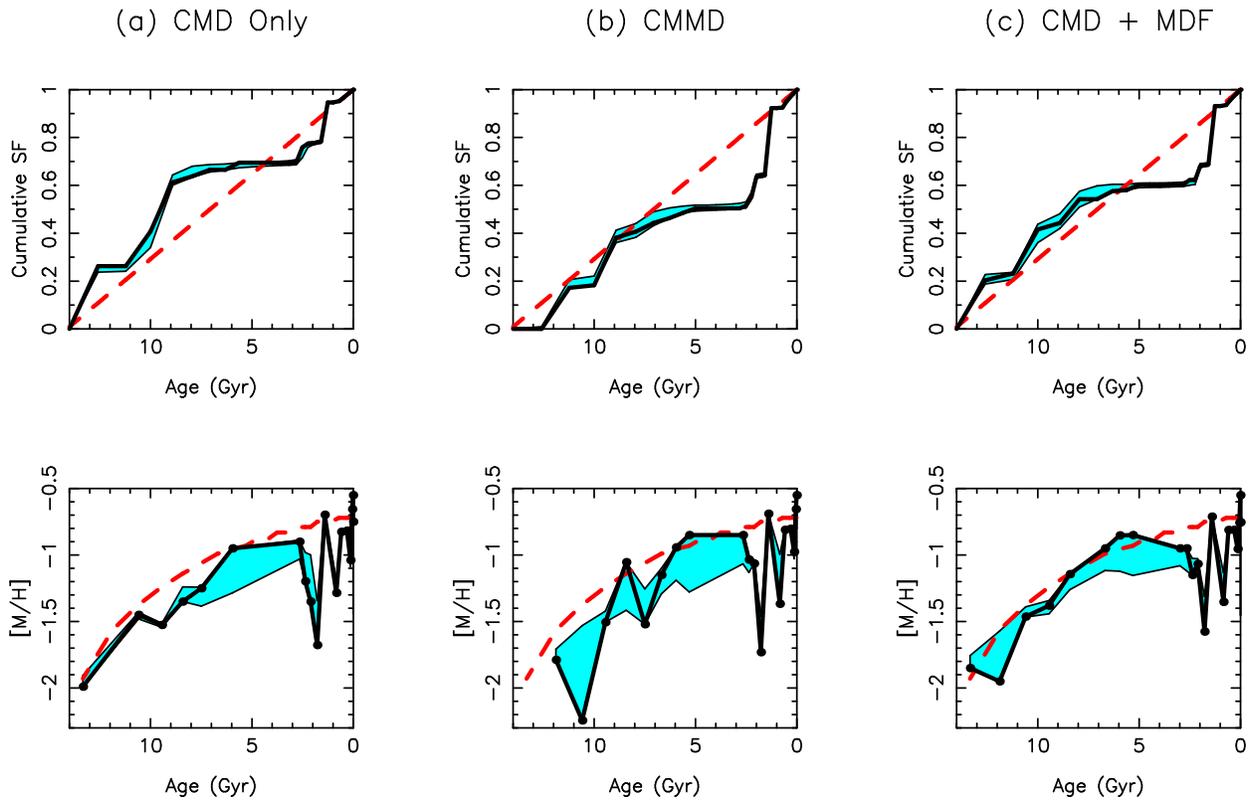}
\caption{Star formation history solutions obtained using the PARSEC models for the data shown in Figure \ref{fig-example2} (which were created using the Padua models).  The three solutions under (a), (b), and (c) are for CMD-only, CMMD, and CMD+MDF solutions respectively.
\label{fig-systematic}}
\end{figure}

\begin{figure}
\epsscale{1.0}
%\plotone{../models/shallow_young_full.bz.ps}
\plotone{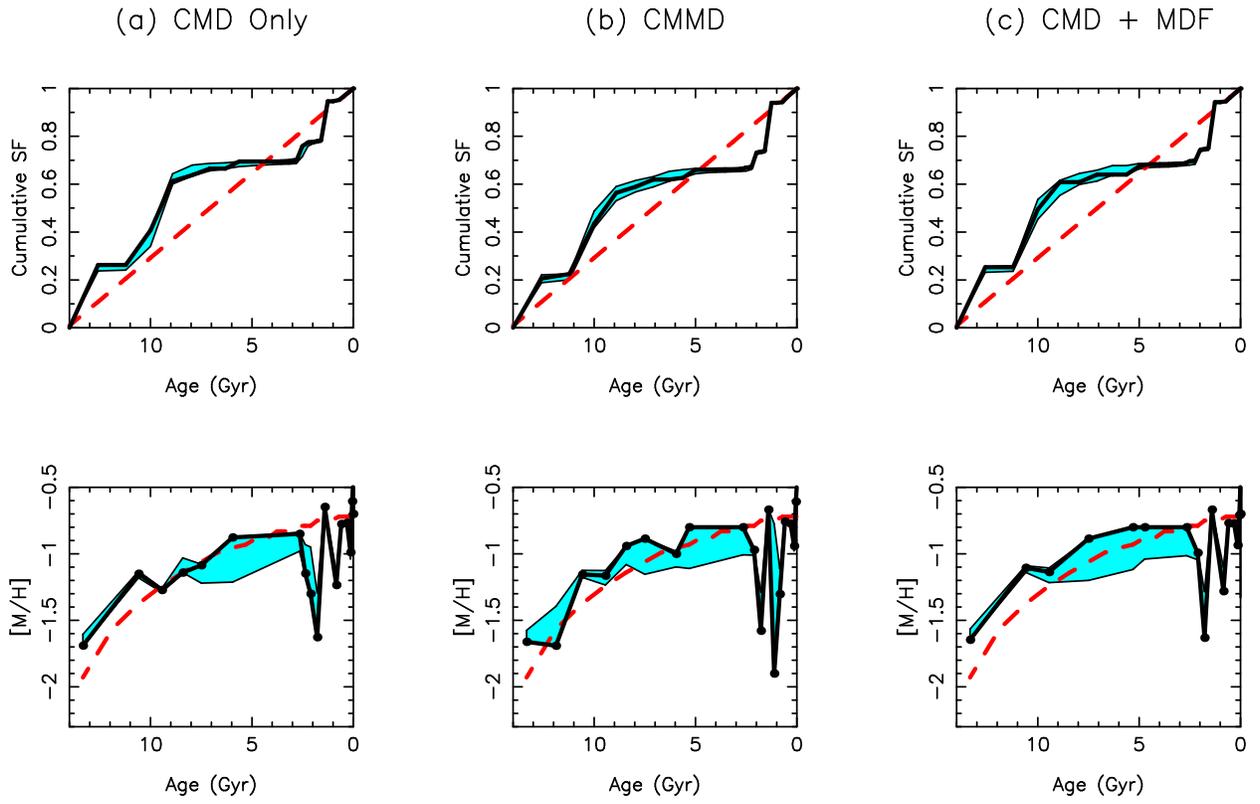}
\caption{The same as Figure \ref{fig-systematic}, but with a metallicity shift applied to better match the CMMD.
\label{fig-sys_fixed}}
\end{figure}

While incorporation of additional data creates a more tightly constrained problem, the results shown in Figure \ref{fig-systematic} show that this can create undesirable results in the particular application of measuring star formation rates.  Specifically, while adding spectroscopic metallicity measurements to photometry creates more precise stellar age estimates, uncertainties in stellar models make these stellar age estimates less accurate than those using photometry only.  Thus, a straightforward incorporation of metallicity measurements into a CMD-fitting code is likely to increase the precision but degrade the accuracy of the measured star formation histories.

An approach for mitigating this error is proposed in Section \ref{sec-correction}.  At its core is a mapping of isochrone metallicities to measured (spectroscopic) metallicities, which can be estimated empirically by finding the mapping that maximizes the goodness of fit.  This is shown to produce SFHs whose SFRs are as accurate as those obtained from CMD-only fits, while the AMRs are consistent with the spectroscopic data.  Note that this mitigation does not eliminate or even reduce systematic errors inherent in CMD-based SFH measurements \citep{dol12}; it only mitigates the increased systematic errors caused by incorporating metallicities into the solutions.  The primary benefit of incorporating spectroscopic metallicity data into SFH solutions is that the resulting AMR is consistent with spectroscopic rather than photometric metallicity measurements.

\acknowledgments

Support for program number HST-GO-13768.05 was provided by NASA through a grant from the Space Telescope Science Institute, which is operated by the Association of Universities for Research in Astronomy, Incorporated, under NASA contract NAS5-26555.

\clearpage

\clearpage

\clearpage

\clearpage

%% The following command ends your manuscript. LaTeX will ignore any text

\end{document}